\documentclass{article}

\usepackage{microtype}
\usepackage{graphicx}
\usepackage{subcaption}
\usepackage{booktabs} 

\usepackage{paralist}
\usepackage[most]{tcolorbox}

\usepackage{hyperref}

\newcommand{\eg}{\hbox{\emph{e.g.,}}\xspace}
\newcommand{\ie}{\hbox{\emph{i.e.,}}\xspace}

\newcommand{\etc}{\hbox{\emph{etc.}}\xspace}
\newcommand{\vs}{\hbox{\emph{vs.}}\xspace}

\usepackage[accepted]{icml2025}

\usepackage{amsmath}
\usepackage{amssymb}
\usepackage{mathtools}
\usepackage{amsthm}
\usepackage{xspace}
\usepackage{multicol}
\usepackage{multirow}
\usepackage{longtable}
\usepackage{inconsolata}
\usepackage{enumitem}
\usepackage{xfrac}
\usepackage{tikz}
\usetikzlibrary{calc}

\usepackage{listings}

\usepackage{textgreek}

\usepackage[capitalize,noabbrev]{cleveref}

\theoremstyle{plain}

\theoremstyle{definition}

\theoremstyle{remark}

\usepackage{soul}
\usepackage{color}

\usepackage{makecell}
\usepackage{pifont}

\newcommand{\probgen}{\textsc{ProbeGen}\xspace}

\DeclareMathOperator{\depth}{depth}

\usepackage[textsize=tiny]{todonotes}

\icmltitlerunning{Disproving Program Equivalence with LLMs}

\begin{document}

\twocolumn[
\icmltitle{Disproving Program Equivalence with LLMs}

\icmlsetsymbol{equal}{*}

\begin{icmlauthorlist}
\icmlauthor{Miltiadis Allamanis}{gdm}
\icmlauthor{Pengcheng Yin}{gdm}
\end{icmlauthorlist}

\icmlaffiliation{gdm}{Google DeepMind}

\icmlcorrespondingauthor{Miltiadis Allamanis}{foo}

\icmlkeywords{llms,code,equivalence}

\vskip 0.3in
]



\printAffiliationsAndNotice{}  

\begin{abstract}
To evaluate large language models (LLMs) for code, research has used manually created unit test-based benchmarks. However, these tests are often inadequate, missing corner cases and other implementation-specific oddities. This work introduces \probgen, a whitebox method that takes two or more executable pieces of code and searches for counterexamples to their equivalence. Comparing code semantics requires a deep understanding of code. We demonstrate that LLMs with execution feedback perform well at this task. In a common code synthesis benchmark, \probgen disproves 18\% of samples considered equivalent to the ground truth by the benchmark-provided unit tests. Additionally, using \probgen, we can semantically cluster LLM samples for semantic self-consistency, improving pass@1 by 10\% by unifying syntactically distinct but semantically similar samples.
\end{abstract}

\section{Introduction}
LLMs have found many successful tasks on coding applications, including code completion, synthesis, editing.
While these methods are often tested and evaluated with execution based feedback, these tests commonly rely on weak tests of equivalence to some ground truth, usually using unit tests~\citep{liu2024your}.
Unit tests commonly perform a limited number of checks for predefined behaviors of the tested code, often missing edge cases or ignoring anomalies that appear within the tested source code.
At the same time, automatically distinguishing semantically non-equivalent versions of code (\eg by providing sample usages where the code’s behavior differs) can be useful to a variety of applications such as explaining code edits~\citep{groninger2024changeguard}, or showcasing how different candidate synthesized code implementations behave~\citep{mayer2015user}.
However, up to today most methods for detecting equivalence rely on formal search or randomized input generation~\citep{fink1997property}.
The recent progress of LLMs’ capabilities in understanding code and its execution~\citep{ding2024semcoder} provides an interesting alternative.

In this work, (\autoref{sec:probe gen}) we are interested in disproving the equivalence of two or more pieces of code by finding a counterexample usage that causes the pieces of code to diverge in behavior, \eg return different values or throw exceptions. Although we will \emph{not} be able to prove the equivalence of any code snippet, a single counterexample suffices to disprove their equivalence. While --- in the general case --- finding a counterexample is undecidable, in practical applications it is often feasible both by knowledgeable humans and LLMs.

\begin{figure}[t]
    \centering
    \includegraphics[width=\columnwidth]{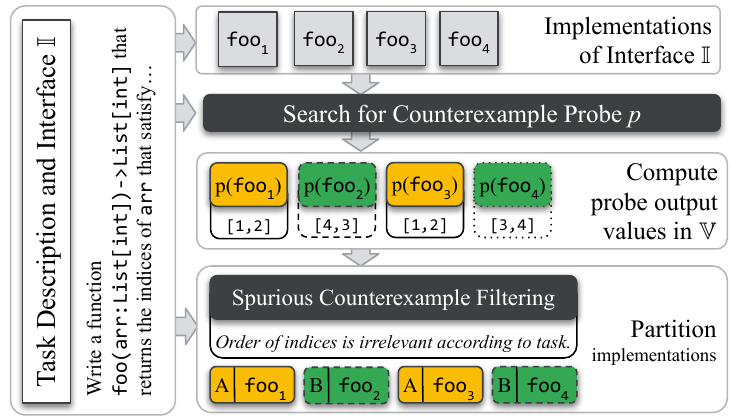}
    \caption{Given a task description, an interface (function \lstinline{foo} in this example) and a number of implementations of the interface, an LLM is asked to generate code (probe) that uses the interface and causes the implementations to diverge in behavior (in this example, return different output lists). This provides a counterexample where the implementations differ (as proven by their execution). However, the -- often implicit -- pre- or postconditions of the interface may be ambiguous. A final step filters spurious counterexamples and yields a partition of the implementations (colored in orange and green) into non-equivalent groups.}
    \label{fig:intro}
\end{figure}

We achieve this through \probgen (\autoref{sec:generating probes with llms}), a search that iteratively invokes an LLM to generate usage examples (probes; \autoref{fig:sample probe}) in a white-box fashion (\ie by looking at the source code). The LLM then receives execution feedback showing the computed probe outputs of the candidate counterexamples. If all outputs match, the LLM proceeds to generate new probes given its ``experience''. When, and if, a counterexample is found it deterministically disproves the equivalence of the candidate code implementations.

However, sometimes, counterexamples violate the (often implicit) preconditions of the task (\eg an input variable representing a counter must always be non-negative) or the outputs differs in unimportant ways (\eg the ordering of the returned items in a list is irrelevant for the task as in \autoref{fig:intro}). To discard these spurious counterexamples, we employ an LLM-based filter (\autoref{subsec:removing fps}).

In our evaluation (\autoref{subsec:search strategy eval}) we present a method for understanding and optimizing the cost-accuracy trade-offs of the counterexample search process in \probgen. Then by drawing a diverse set of code samples for a set of tasks, in \autoref{subsec:probes for synthesis} we show that pre-existing unit tests sometimes fail to capture important semantic differences among code implementations of the same task (in our dataset 24\% of samples that pass unit tests, are disproved to be equivalent to the ground truth, and this figure goes to 18.7\% if we account for errors in the spurious counterexample filtering) showing that \probgen can either be used to complement existing test suites or provide an alternative to unit testing when tests do not already exist.

Finally, we show that we can use \probgen to perform semantic clustering (\autoref{sec:semantic clustering}), \ie partition a set of implementations based on their semantics, allowing to more easily perform error analyses and in \autoref{subsec:ssc} we show that when used as a novel form of self-consistency it offers mild performance improvements improving pass@1 by about 10\%. 

\paragraph{Contributions} In summary our contributions are:
(a) Introduce the problem of disproving functional equivalence of a program with probes as a machine learning task that requires deep program understanding and has multiple applications;
(b) Show that probe generation is a useful and effective methodology for evaluating synthesized code.
This approach complements unit tests and reduces the need to create ones.
(c) We show that LLMs enable a white-box, feedback-driven search process for LLMs to address the task.
(d) Present and evaluate a novel semantic clustering method for performing error analyses and improving LLM outputs via semantic self-consistency.

\section{Disproving Functional Equivalence via Probe Generation}
\label{sec:probe gen}
Given an interface (API) $\mathbb{I}$, and two or more implementations $f_1, f_2, ... : \mathbb{I}$ of the interface,
our goal is to disprove the functional equivalence of these implementations.
The interface $\mathbb{I}$ might be a function signature, a class-like interface, a package, a protocol, \etc
Although, it is well-known that determining the equivalence of two programs is generally undecidable, for a large subset of real-life programs disproving functional equivalence is possible by humans and --- as we will show soon --- by LLMs alike.

\begin{figure}[t]
\centering
\begin{subfigure}{0.49\columnwidth}
\begin{lstlisting}[basicstyle=\ttfamily\scriptsize, language=Python]
def probe():
 a = Clazz(1,"sample")
 b = a.foo(0)
 c = b.bar(a.baz("3"))
 return set(b - c)
\end{lstlisting}
\caption{Probe}\label{fig:sample probe}
\end{subfigure}
\begin{subfigure}{0.5\columnwidth}
\begin{lstlisting}[basicstyle=\ttfamily\scriptsize, language=Python]
def test():
 a = Clazz(1,"sample")
 b = a.foo(0)
 c = b.bar(a.baz("3"))
 v = set(b - c)
 assert len(v) == 7
\end{lstlisting}
\caption{Unit Test}\label{fig:sample unit test}
\end{subfigure}
\caption{Sample probe \vs unit test for a fictional class interface \lstinline{Clazz}. 
A probe only provides inputs in a usage example, whereas a unit test needs to make concrete assertions about the expected values.
This makes probes easier to generate and allows us to perform more accurate testing for equivalence.}
\label{fig:probes vs unit tests}
\end{figure}

While there are many ways to disprove functional equivalence, we formulate the task as a probe generation task, which can be thought as a form of differential testing~\citep{mckeeman1998differential} using the ``equivalence modulo inputs'' paradigm of \citet{le2014compiler}.
Specifically, we call a \emph{probe function}, a function $p_i: \mathbb{I} \rightarrow \mathbb{V}$ that accepts an implementation of the interface $\mathbb{I}$ and returns a value in $\mathbb{V}$, where $\mathbb{V}$ is the set of values where an (in)equality operator can be defined and computed in a bounded amount of time (see \autoref{fig:sample probe} for an example).
If $\exists i,j \vert i\neq j \land p_k(f_i) \neq p_k(f_j)$, then the probe $p_k$ exposes a counterexample showing that $f_i$ and $f_j$ are \emph{not} functionally equivalent.
Note that this formulation assumes that (a) there is no randomness involved in computing $p_k(f_i)$, (b) all (if any) side-effects can be captured by a probe $p_k$ and (c) $p_k$ tests only the functional equivalence rather than some non-functional property (\eg memory usage) and (d) $p_k$ does not probe the internals of its input implementation (\eg via reflection).
Note that commonly-used unit tests are a special limited case of probes where $\mathbb{V}=\{\textsf{True}, \textsf{False}\}$, returning \textsf{True} when a test passes, and \textsf{False} when a test fails.
Unit tests achieve this by comparing expected values with the actual values returned (\autoref{fig:sample unit test}).
In contrast to that, probes need to only provide inputs to $\mathbb{I}$ and do \emph{not} need to make any assertions about the expected output values (\autoref{fig:probes vs unit tests}).
This difference makes generating probes an easier and more general task compare to generating unit tests since only valid inputs need to be generated.

\section{Generating Probes with LLMs}
\label{sec:generating probes with llms}
We now focus on the task of probe generation with LLMs.
LLMs have had a wide success with a number of coding tasks such as code generation and repair.
There are two important aspects that make LLMs attractive for the task: (a) given their ability to inspect the source code, they can be used as a white-box probe generation method.
While this requires advanced code understanding capabilities, LLMs have shown an increasing ability to understand code's complex structure and execution behavior.
(b) LLMs have shown increasing abilities to take into account execution feedback, \ie information about the concrete execution of code, such as outputs and error messages.~\citep{chen2023teaching}

Given these, we formulate the task --- which we call \probgen --- as a multi-turn generation problem where an LLM accepts as input two or more implementations $\{f_i\}$ of $\mathbb{I}$ and the list of previously generated probes $p_{\tau}$ and their output values $p_{\tau}(f_i)$ of each probe for every implementation $f_i$. The LLM is then asked to sample the next probe, \ie 
\begin{align}\label{eq:probe generation}
p_t \sim  \textsf{LLM}\left(\{f_i\}, \bigcup_{\tau=1}^{t-1}\left\langle p_{\tau}, p_{\tau}(f_1), p_{\tau}(f_2), ...\right\rangle\right),
\end{align}
where $p_{\tau}$ is the probe sampled at the $\tau$-th turn, $\{f_i\}$ is the set of implementations considered, $\left\langle p_{\tau}, p_{\tau}(f_1), p_{\tau}(f_2), ...\right\rangle$ is the tuple of the probe generated at time $\tau$ along with the (string-represented) values $p_{\tau}(f_i)$ returned by the probe for each $f_i$.
Essentially, \probgen performs guided differential testing using both execution feedback and the source code of the implementations.
\autoref{appx:samples} shows two examples (that are too long to fit here) where \probgen generates a probe that exposes a counterexample, but the preexisting unit tests deem the two implementations to be equivalent.

While the LLM views \probgen as a sequential, iterative process, we can use \probgen as a tree search where branching occurs by sampling the LLM multiple times with a given context using \autoref{eq:probe generation}.
In \autoref{sec:evaluation} we use \probgen in a tree search process sampling $k$ samples for each context.
In \autoref{subsec:search strategy eval} we are going to evaluate different search strategies for generating probes and discuss the tradeoffs involved in branching \vs increasing the number of turns.

Note that if we want to reuse the probes for other implementations of $\mathbb{I}$, we can pack all the probes into $p_{\text{comb}}=p_{1} \circ p_2 \circ \dots$ that combines the output of all probes into a tuple.
Alternatively, if we are just interested in disproving the functional equivalence of $f_i$ and $f_j$ (\eg as in \autoref{sec:evaluation} for evaluating program generation) using only the probe(s) where $p_k(f_i)\neq p_k(f_j)$ suffices.

\paragraph{Limitations}
It should be acknowledged that \probgen has some limitations:
\begin{inparaitem}
    \item Code must be executable and deterministic on the explicit inputs provided.
    \item The value equality operator must be provided and probes must return readily comparable values, avoid infinite generators, functions that return functions, \etc.
    \item Generating probes with an LLM is much more costly compared to running unit tests since querying an LLM is much more costly that using a few CPU-minutes. In that sense, this approach makes sense only when unit tests --- if they exist --- have already failed to disprove equivalence.
    \item Side-effects may go unnoticed, unless there are rigorous checks and sandboxing.
\end{inparaitem}

\subsection{Filtering Spurious Counterexamples}
\label{subsec:removing fps}
Counterexamples exposed by probes fully disprove the strict functional equivalence. 
However, there are cases where the (implicit) preconditions of the interface $\mathbb{I}$ are violated or the differences in the output are superficial and not covered by the provided equality operator.
For example, consider two implementations of a function \texttt{isPrime(n)} whose behavior differ when passing floats, negative integers (\eg one throws an exception whereas the other return \texttt{False}).
There are also cases where the output values are different, but in a way that does not matter to the problem definition.
For example, a task may ask to return the list of indices of an array that satisfy some property (\autoref{fig:intro}).
If the the task does not specify the order of the returned indices, the returned values \lstinline{[3,4]} and \lstinline{[4,3]} should be considered equivalent despite the fact that the two returned lists differ.

A definition of what is or is not a spurious counterexample generally depends on the target use case.
Sample 2 in \autoref{appx:samples} shows such an example:
\probgen finds a corner-case for two implementations of \lstinline{count_string()}.
The task asks for code that counts the number of case-insensitive occurrences of some string within some text.
In these particular implementations, \probgen finds a string containing Greek characters that causes the two implementations to return different values\footnote{In particular one implementation first lower-cases the input string, causing ``\textSigma\textSigma'' to be lower-cased as ``\textsigma\textvarsigma'' as defined in the Unicode standard, whereas the other implementation lower-cases each character independently lowercasing ``\textSigma\textSigma'' to ``\textsigma\textsigma''.}.
Such a semantic difference might be irrelevant when treating the task as a student exercise, but could be significant if the code was intended to be used in production across multiple locales.

In the case of natural language specifications and in this work, we consider a spurious counterexample as one exposing a difference due to inherent ambiguities in natural language or the expected behavior is left unspecified.
Filtering spurious counterexamples can be performed by prompting an LLM at the risk of introducing false negatives.
The method of \citet{endres2023formalizing} who use task specifications to generate a set of preconditions (formal conditions) could complement our filtering by detecting inputs that violate the preconditions.
We expect that the method of \citet{endres2023formalizing} would offer different precision-recall trade offs, but we do not consider it in this work given its complexity.

\subsection{Applications}
\probgen may find multiple applications in software engineering and the domain of AI for code.
One application --- discussed in \autoref{sec:evaluation} --- is to evaluate code synthesis and editing tasks where a ground-truth solution is available but the unit tests are weak or non-existent.
As we show in \autoref{sec:evaluation} when unit tests are not available but a ground-truth implementation is, \probgen can behave similarly to unit testing or can complement weak test suites.  

Another application is the task of explaining (to a human or an LLM) the semantic differences between two snippets of code by providing showing probes and their outputs.
For example, \probgen can find probes that differentiate some code before/after some edit, expose differences introduced during refactoring~\citep{groninger2024changeguard}, explain mistakes in student code given an existing solution, \etc

Finally, \probgen can be used to perform semantic clustering (\autoref{sec:semantic clustering}) when multiple implementations $\{f_i\}$ are available.
Semantic clustering can be used for performing semantic self-consistency, \ie finding the largest semantic cluster among multiple candidates.
We evaluate this in \autoref{subsec:ssc}.
Semantic clustering can also find application in the explainability of LLM suggestions, including error analysis, and in human-computer interaction of code synthesis and editing systems where users are presented with multiple alternatives along with concrete examples of their semantic differences, generalizing the work of \citet{mayer2015user}.

\section{Evaluation}
\label{sec:evaluation}
In this section we evaluate \probgen as an evaluation method for code generation tasks, such as code synthesis and repair, and contrast it to other (non-LLM-based) methods.
Our goal here is \emph{not} to show which LLM performs best in this task\footnote{Comparative results among LLMs would be ephemeral and likely to have changed between the time the reader started reading this paper and finishing reading this footnote.}, but to establish \probgen as a viable and strong method for disproving functional program equivalence.

\paragraph{Experimental Setup}
We implement \probgen (\autoref{eq:probe generation}) by specializing it to (a) accept only two implementations $f_1, f_2$ at each time. This reduces the complexity of the problem and allows the LLM to focus on the differences of the two implementations.
(b) we focus on Python, fixing the execution environment to common libraries used in our code generation benchmark;
(c) we ask the LLM to generate a set of arguments and keyword arguments that will be input to the function interface being tested.
(d) we ask the LLM to create a Python generator that yields one or more inputs at each turn.
This allows the LLM to use arbitrary code (\eg \texttt{for} loops) to generate multiple inputs/probes at each turn.
(e) we type check (using use the \texttt{typeguard} package) to ensure that the inputs provided match the interface's type annotations (if any).
This check avoids generating counterexamples that violate the (explicit) preconditions set by the function signature.
Note that this implementation of \probgen is limited to testing function interfaces, but allows us to ensure that the generated inputs are passed into the implementations efficiently and reduces the complexity of spurious counterexample filtering.
We filter spurious counterexamples using the method of \autoref{subsec:removing fps}, the prompt of \autoref{appx:fp filtering prompt}, Gemini Flash 2.0\footnote{\href{https://deepmind.google/technologies/gemini/flash/}{deepmind.google/technologies/gemini/flash/}}, and greedy decoding.

We implement the value (in)equality operator ($\neq$) special-casing particular comparisons and delegating the rest to Python's default \texttt{\_\_eq\_\_} implementation.
Specifically, for \texttt{float}s we consider two values to be equal if they are close (using the default \texttt{math.isclose} parameters) but also ensure that two \texttt{NaN} values are considered equivalent.
For NumPy arrays and Pandas frames, we use the relevant methods provided (since the overloaded \texttt{==} operator performs a different function).
If a function returns an iterable we resort to checking only their first 1k elements.
Finally, we consider all exceptions equivalent independent of their type or error message.
While the above design decisions are reasonable for our current experimental setup different implementations may be needed for other tasks or domains.

\paragraph{Evaluation Data}
We focus on the Python LBPP benchmark~\citep{matton2024leakage} which is a code synthesis benchmark that is relatively new and unleaked.
The LBPP dataset provides a set of unit tests and a ground-truth implementation.
We query a variety of models to generate predictions for the LBPP dataset using the prompts shown in \autoref{appx:lbpp implementations}.
The goal of the sample generation is \emph{not} to maximize correctness.
Instead our prompt aims to get diverse solutions, possibly with a wide variety of semantic mistakes and differences.
Our goal is to compare those implementations to the ground truth and show the differences among \probgen and other methods.
We generate about 10k samples in total.

\subsection{Evaluating Search Strategies}
\label{subsec:search strategy eval}

\begin{figure}[tb]
    \centering
    \includegraphics[width=.8\columnwidth]{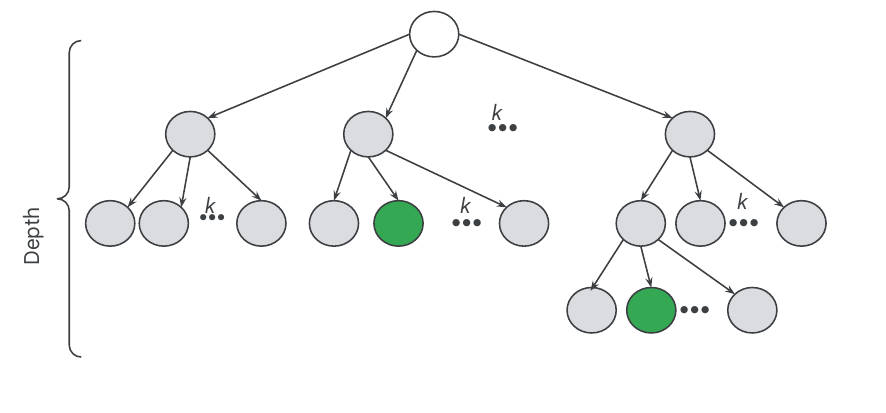}
    \caption{\probgen as a tree search process. At each step $k$ independent samples may be drawn, up to a maximum depth $D$. When a probe successfully disproves the equivalence (green nodes) it forms a leaf node.}
    \label{fig:search tree}
\end{figure}
We now focus on showing the trade-offs involved in the probe generation tree search process.
In particular, we are interested in the tradeoff between taking more independent samples \vs multiple turns of feedback and probe generation (\autoref{fig:search tree}). 
We limit ourselves to a single model --- Gemini Flash 2.0 --- given its cost-performance trade-offs and the costly nature of this particular experiment: a relatively cheap/small LLM is more appropriate when \probgen is used for evaluation.
While it is reasonable to expect that the exact results will differ across models and their versions, the methodology is general and can be used to tune the probe generation process for an arbitrary LLM.

\paragraph{Probability of Success for Tree-search processes}
To evaluate different search strategies we need to measure their effectiveness.
In particular, we are interested in computing the probability that a particular search strategy strategy will be able to disprove the functional equivalence of two or more implementations assuming that one exists.
To measure this, we generalize the commonly used pass@$k$ estimator~\citep{chen2021evaluating} to tree search processes.
In particular, given a search strategy $\mathcal{S}(d)$ which returns the branching factor $k$ for the search at depth $d$ (\autoref{fig:search tree}), the probability of success $\sigma$ at node $n_i$ can be recursively defined as
\begin{equation*}
    \sigma(n_i, \mathcal{S}) = \left\{
        \begin{array}{l}
        1 \text{,~if~} \exists l, m: p^{(n_i)}(f_l) \neq p^{(n_i)}(f_m) \\
        \frac{\sum_{\mathbf{c} \in C_\mathcal{S}(n_i)}\left(1 - \prod_{n_j\in \mathbf{c}}\left(1-\sigma(n_j)\right)\right)}{|C_\mathcal{S}(n_i)|} \text{~, otherwise} \\
        \end{array}
    \right.
\end{equation*}
where $C_\mathcal{S}(n_i)$ is the set of all $\mathcal{S}(\depth(n_i))$-sized combinations of the children of $n_i$.
This estimator conceptually performs the same operation as pass@$k$ recursively at each node of the search tree but accepts probabilities rather than just 0/1 (fail/pass) events.

\begin{figure*}[t]
    \centering
    \begin{subfigure}{0.31\textwidth}
        \centering
        \includegraphics[width=\textwidth,trim={1cm 0.1cm 1cm 1cm},clip]{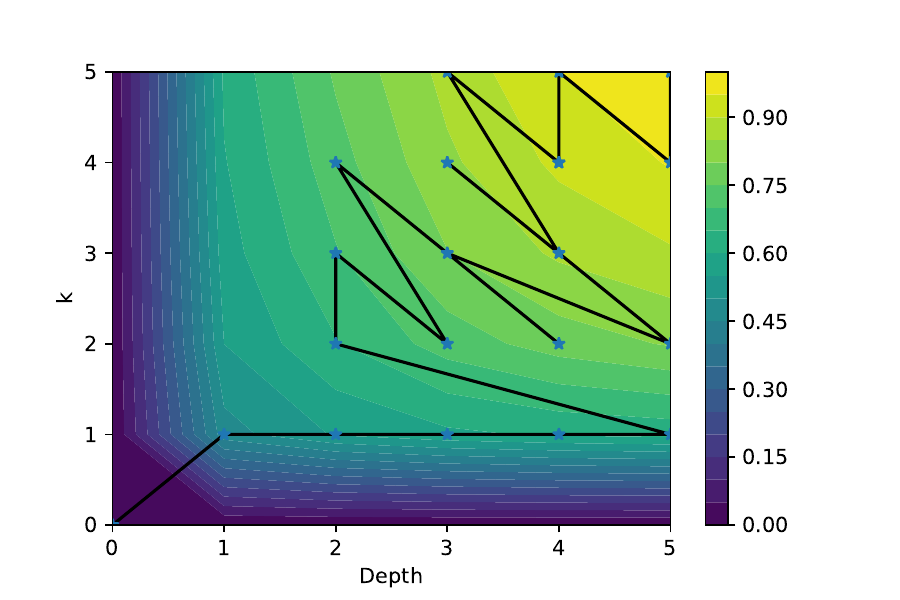}
        \caption{\textsf{Full} Search Strategy}\label{fig:search tree full}
    \end{subfigure}
    \begin{subfigure}{0.31\textwidth}
        \centering
        \includegraphics[width=\textwidth,trim={1cm 0.1cm 1cm 0.5cm},clip]{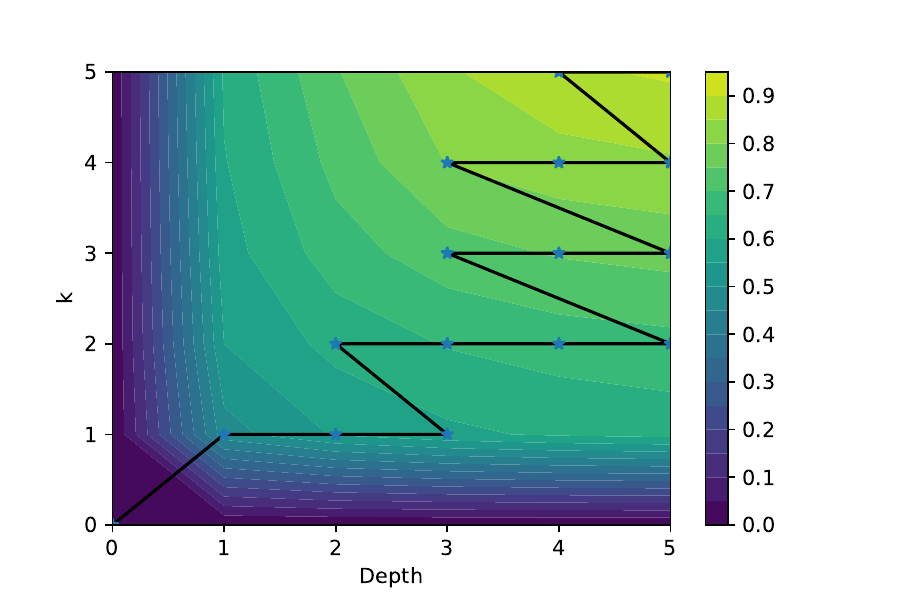}
        \caption{\textsf{Decreasing} Search Strategy}\label{fig:search tree dec}
    \end{subfigure}
    \begin{subfigure}{0.37\textwidth}
        \centering
        \includegraphics[width=\textwidth,trim={0.1cm 0.1cm 1cm 0.5cm},clip]{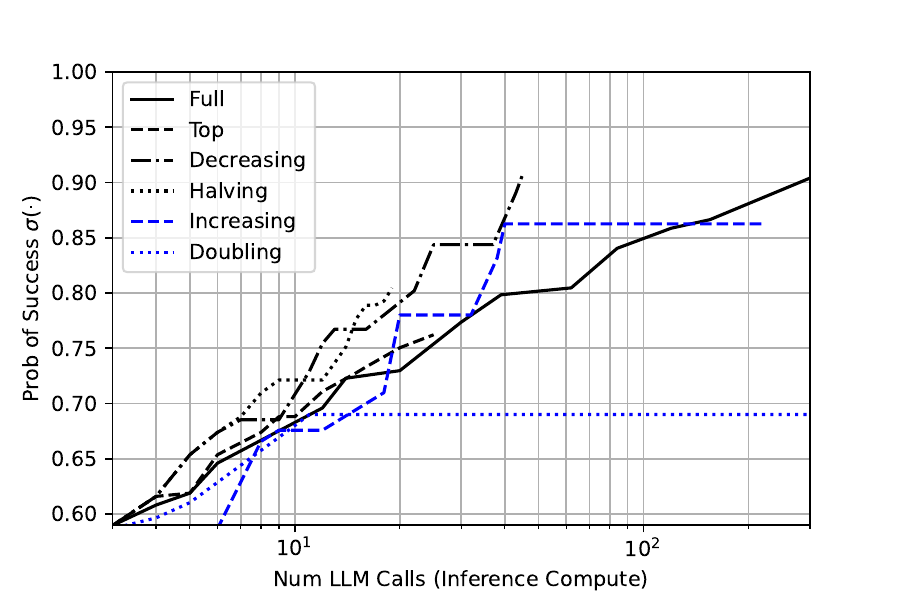}
        \caption{Best probability of success $\sigma(n_\textsf{root}, S)$ \vs inference cost for each search strategy. Note $\log x$ axis.}\label{fig:strategy comparison}
    \end{subfigure}
    \caption{Relationship between the branching factor $k$ and search depth to the probability $S$ of successfully generating a probe that exposes a counterexample and the trade-offs with the inference cost. The two contour plots (left) show the probability of success (lighter color is higher) with respect to $K$ and depth for two search strategies (plots for all strategies can be found at \autoref{fig:search trees all}). The lines in the contour plots trace (from the bottom left) the Pareto-optimal trade-off between inference cost and probability of success $S$. \autoref{fig:strategy comparison} plots the best probability of success $S$ for a given inference cost for each search strategy.}
    \label{fig:search trees}
\end{figure*}

To measure this, we sample from our evaluation data 600 LBPP generated implementations, 300 of which pass the LBPP unit tests and 300 fail them. We then sample the full search tree with a branching factor $K=5$ and up to a depth of $D=5$.
This is a costly process since a full tree of depth $D$ and branching $K$ has $(K^{D+1}-1)/(K-1)-1$ non-root nodes, \ie LLM invocations.
Note the search tree might not be full when some node at depth $d<D$ disproves the equivalence of the implementations and further turns make no sense.
This process yields 449 search trees where at least one node generated a probe that can be used to disprove the equivalence of the implementation to the ground-truth LBPP implementation.
We then consider the following search strategies representing different tradeoffs between branching $K$ and depth $D$. 
\vspace{-1em}
\begin{itemize}[leftmargin=*,noitemsep,parsep=0pt]
    \item \textsf{Full}: Constant branching $K$ up to depth $D$.
    \item \textsf{Top}: Sample $K$ independent trajectories up to depth $D$.
    \item \textsf{Decreasing}: Sample $k=\max(1, K)$ times at the top level, $\max(1, K-1)$ in the next level, \etc up until depth $D$.
    \item \textsf{Increasing}: The same as above but incrementing $K$ instead.
    \item \textsf{Halving}/\textsf{Doubling}: The same as above but halving/doubling $k$ at each turn.
\end{itemize}
\autoref{fig:search tree full} and \autoref{fig:search tree dec} plot the contour plots for different values of the parameters $K$ and $D$ of two search strategies.
Plots for all strategies are shown in \autoref{fig:search trees all} in the appendix.
In addition to the performance for a given $D$ and $K$, the figures show the cost-to-performance Pareto front, where cost is measured by the number of LLM calls required --- a reasonable proxy to the actual inference cost.
\autoref{fig:strategy comparison} takes the Pareto front for each search strategy and plots the compute cost \vs the probability $\sigma$ of succeeding in finding a disproving probe.

As \autoref{fig:search trees} shows, the probability of finding a disproving counterexample increases as the compute spent increases.
Deeper search (multiple turns) are favored compared to more samples, but some sampling also provides useful diversity and avoids situations where the LLM gets ``stuck'' in a particular mode.
The \textsf{Decreasing} search strategy (\autoref{fig:search tree dec}) stands out as a particularly competitive strategy that trades-off initial sample diversity with increased depth: \autoref{fig:strategy comparison} shows it can achieve 90\% probability of success with about 4x less cost that the \textsf{Full} search strategy.
The \textsf{Increasing} strategy is also competitive for a mild number of LLM turns, whereas the commonly used \textsf{Top} strategy performs about as well as \textsf{Full}.

While the above results may vary for different models the methodology presented here is general and can be used to pick an optimal cost-performance point.
Future work may also investigate more complex, learned search strategies.

\subsection{Probe Generation for Code Synthesis Evaluation}
\label{subsec:probes for synthesis}
Given the above and \autoref{fig:strategy comparison} we opt for using the \textsf{Decreasing} strategy with an initial branching factor $K=3$ and maximum depth $D=4$.
Thus the maximum number of LLM calls for a single sample is 21 -- significantly less than the per-sample cost required in the previous subsection.

\paragraph{Baselines}
Since \probgen can be used to evaluate code synthesis, we evaluate two other baselines.
First, unit tests are a common form of evaluating synthesized code.
While unit tests do not require a ground-truth solution, they take a black-box approach to the tested code potentially missing important edge cases or ignoring ``oddities'' present in the code.
At the extreme, a code implementation that merely uses a dictionary to map all the input values in a test suite to the output values would be considered correct by the unit tests but would fail to actually solve the target task.

An alternative to unit testing is property-based testing.
Property-based testing (PBT) can be thought as a form of fuzzing or random testing, where inputs are randomly generated (discarding any inputs that don't meet any preconditions), passed to the code-under-test, and finally the outputs are checked for a desired property~\citep{fink1997property}.
Random differential testing is one form of property-based testing.
In this work, we use the Hypothesis\footnote{\url{https://hypothesis.readthedocs.io/}} Python package for performing equivalence testing of functions.
In contrast to \probgen, property-based testing takes a black-box approach to testing, but in the limit guarantees that functional equivalence can be disproved.
One advantage of this approach compared to \probgen is that it requires CPU-time rather than specialized hardware that LLMs commonly require.
In this work, we use a timeout of 3 CPU-minutes per target.
Similar to our approach, PBT does not require pre-existing unit tests.
The main disadvantage of PBT is that it requires (a) rigourous pre-conditions for the random inputs generated and (b) a good random input generator, which is often hard to construct for structured data.
In this evaluation, we use the type annotations (if present) as preconditions and rely on the Hypothesis' built-in input generators.

\paragraph{Results}
\autoref{tbl:quality evals} shows the proportion of samples runnable by a pair of methods along with the percent of samples that each method has (dis)proved their equivalence.
First, we can see that not all methods can actually run the code samples.
Unit tests run on all examples by construction of the LBPP dataset, although in many real scenarios they might not be available without significant human effort.
PBT (and the Hypothesis) package has the smallest percent of runnable samples. 
This is because Hypothesis requires valid type annotations in the input arguments for the generation of random inputs, and for the input types to be within a known set (custom random value generators for new types are possible, but out of scope).
Finally, \probgen cannot run 11\% of the samples.
This is because of a variety of factors: (a) some samples cannot be tested with our functional probe generation (\eg function decorators) (b) the LLM failed to generate inputs that comply with the type annotations (c) the task cannot be tested by providing inputs to a function (\eg the test target is a class or a function decorator).

\newcommand{\wrongMark}{{\Large\textcolor{red}{\pmb\times}}}
\begin{table*}
\centering
\caption{(Dis)agreement between pairs of methods over the samples that both methods can run. For a given sample, a method may consider it correct ($\checkmark$) or incorrect ($\wrongMark$). }\label{tbl:quality evals}
\begin{tabular}{llrrrrr} \toprule
     &  & \multicolumn{4}{c}{\% among mutually runnable samples} & \multirow{2}{3cm}{\% Mutually Runnable Samples} \\ \cmidrule{3-6}
     && $(\checkmark, \checkmark)$ & $(\checkmark, \wrongMark)$  & $(\wrongMark, \checkmark)$ &  $(\wrongMark, \wrongMark)$ &  \\ \midrule
Unit Tests &\vs PBT & 10.8 & 28.2 & 7.6 & 53.4 & 59.6 \\
Unit Tests &\vs \probgen & 33.6  & 10.8  & 5.9 & 49.7 & 89.0 \\
Unit Tests &\vs \probgen w/o Filter & 25.2 & 19.1  & 3.4 & 52.2 & 89.0 \\
PBT &\vs \probgen & 9.3 & 7.9 & 22.5 & 60.4 & 54.8 \\
PBT &\vs \probgen w/o Filter & 6.8 & 10.3 & 13.7 & 69.2 & 54.8 \\
\probgen &\vs \probgen w/o Filter & 28.7 & 10.8 & 0.0 & 60.6 & 89.0 \\ \bottomrule
\end{tabular}
\end{table*}

\autoref{tbl:quality evals} shows a substantial number of samples where one method finds a sample to be equivalent to the ground-truth but another method finds a counterexample to disprove their equivalence.
While some portion of those counterexamples may be spurious (as defined in \autoref{subsec:removing fps}), this result highlights the diversity of semantic differences among various code samples drawn from an LLM that may (or may not) be of importance to the users of those samples.
In particular, \probgen finds 24.3\% of the samples that pass the unit tests (which consist of 44.4\% of all the samples) to have a significant semantic difference to the ground truth.
This shows the shortcomings of unit tests, which can sometime fail to predict and capture ``oddities'' appearing in generated code, that the white-box approach of \probgen can often detect.
If we were to remove the spurious counterexample filtering, an additional 18.8\% of the examples would be deemed to be different from the ground truth.
PBT --- and Hypothesis in particular --- suffer from many spurious counterexamples, due to the lack of strong preconditions (only the type annotations are available).
This shows the prevalence of implicit pre- and postconditions of the interfaces in the LBPP benchmark and the wide range of 
behaviors code can have when the code is operated outside of the intended setting.

At the same time, 12.5\% of the samples where \probgen failed to generate any counterexamples within the allocated compute budget, the unit tests disproved the equivalence of the sample to the ground truth suggesting that --- as one would expect --- with limited budget \probgen can also fail to disprove the equivalence of two snippets of code.
It should be noted that \probgen and unit tests agree on 83.3\% of the outcomes showing that these two methods agree quite often and that \probgen can constitute a good alternative to unit tests when those do not exist.
However, note that given the relatively costly nature of \probgen compared to unit tests, if unit tests exist, it makes sense to only run \probgen on samples where all unit tests pass.

\paragraph{Human Evaluation}
We sample 2 samples per LBPP problem (if more than 2 exist) that are considered correct by the unit tests but \probgen exposes a counterexample.
Two authors of this work annotate whether the counterexample is valid given the problem definition.
The annotators only view the problem definition, the function signature, inputs and outputs to determine whether the difference in outputs is indeed significant given the natural language specification with instructions to follow the prompt in \autoref{appx:fp filtering prompt}.
This information is identical to the one provided to the LLM filter (\autoref{subsec:removing fps}).
Over the subset of examples that were annotated by both annotators the inter-annotator agreement is ``substantial'' ($\kappa=0.71$), but not perfect.
Upon reviewing the cases of disagreement, it is clear that some edge cases actually constitute a spurious counterexample, is contentious (\eg Sample 2 of \autoref{appx:samples}).
Overall, the precision of the spurious counterexample filtering is 77.1\% according to the human evaluation.
Using this figure, we can estimate that 18.7\% of the samples that pass the unit tests actually have a significant semantic difference to the ground truth.
The filter seems to have some common failure cases when verifying that the input meets the preconditions is hard (\eg an input graph does not have more than one circle), some string manipulation, and cases where the problem definition is too ambiguous regarding the input preconditions.

\subsection{Qualitative Analysis for \probgen}
Finally,we inspect the search trees towards gaining useful insights.
First, we observe that for most samples where both unit tests fail and \probgen finds a counterexample, the search process is relatively short and \probgen could find a counterexample early in the search.
A large set of spurious \probgen counterexamples that are later filtered out are cases where the implicit preconditions are violated and usually \probgen needs to perform deep search, receiving multiple rounds of execution feedback before looking for these edge cases.
For example, for a task that requires passing the number of students, \probgen finds counterexamples when invoking the function with a negative number of students.
The handling of such edge cases is often quite different across implementations: some implementations might explicitly raise an \texttt{Exception} but others may return a nonsensical value.
While it is reasonable to assume that such differences should be ignored for evaluating program synthesis capabilities (as our filter does), the differences in these behaviors might cause or propagate software bugs in real-life software systems.
In most cases, \probgen filters such cases as spurious.
A second common set of spurious counterexamples filtered out are cases where the output type is not well-defined in the problem definition or the output type's default equality implementation does \emph{not} match the semantic notion of the equality required by the problem definition.
For example, one task requires returning a list of indices that have a certain property (\autoref{fig:intro}).
However, the order of the elements of the list is not specified and different implementations return the indices in a different order.
While these lists are different, the task implicitly requires comparing the sets of indices rather than the lists.
Additionally, inspecting some of the cases where unit tests pass, but \probgen finds a counterexample we find two groups: problem definitions with weak test suites that fail to test important aspects of the functionality, or probes that test implementation-specific edge cases.

Finally, we found one case where the LLM tried to disable \texttt{typeguard} dynamically before generating inputs.
While it failed to achieve this, this example shows the importance of setting up guardrails and sandboxing to avoid ``cheating''.

\section{Semantic Clustering}
\label{sec:semantic clustering}

\begin{figure}[t]
    \centering
    \includegraphics[width=\columnwidth]{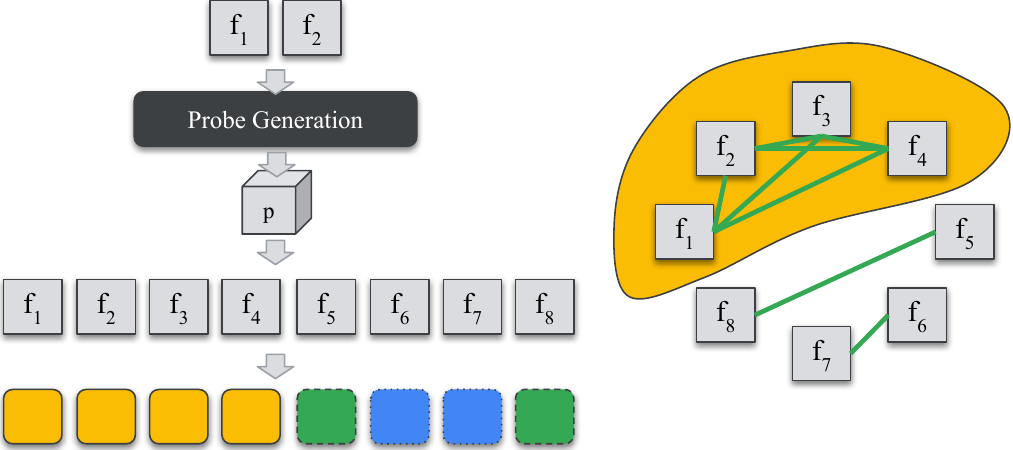}
    \caption{Semantic Clustering through Probe Generation. Given a number of implementation $f_1, ...$ of an interface $\mathbb{I}$ we can generate probes that partition the set of implementations based on their outputs, clustering the implementations based on their semantics.}
    \label{fig:semantic clustering fig}
\end{figure}

Given a probe-generating process, such as the one described in \autoref{sec:generating probes with llms}, we can use it to perform semantic clustering (\autoref{fig:semantic clustering fig}).
Specifically, given $K$ implementations $\{f_i\}$ of an interface $\mathbb{I}$ and a probe function generator we can attempt to partition $\{f_i\}$.
This can be achieved by generating a series of probe functions $p_1, p_2, ...$ and ensuring that if $\exists k: p_k(f_i) \neq p_k(f_j)$ then $f_i$ and $f_j$ appear in separate clusters.

We implement semantic clustering using \probgen (\autoref{sec:generating probes with llms}) as an iterative process.
We start by placing all $\{f_i\}$ into a single cluster.
At each step we randomly pick a random pair of implementations $f_j$ and $f_k$ from the largest cluster, ensuring that the pair has not been previously picked.
We then use \probgen to generate a probe $p_{j,k}$ and compute $p_{j,k}(f_i)$ for all $\{f_i\}$.
We split any clusters whose elements are discriminated by a counterexample which is deemed as ``not spurious'' (using the filter described in \autoref{subsec:removing fps}).
While we could exhaustively check all possible pairs, we limit our method to $N_P$ iterations, generating $N_P$ probes.

\subsection{Semantic Self-Consistency}\label{subsec:ssc}
Semantic clustering enables a novel \emph{semantic self-consistency} (SSC) --- a form of self-consistency~\citep{wang2022self,chen2023universal} for code-like tasks.
While ``traditional'' self-consistency would require a very restrictive notion of equality among candidate solutions (\eg token-level equality), semantic self-consistency uses a less restrictive approach allowing semantically equivalent but syntactically different solutions to be considered equivalent while maintaining the essence of self-consistency.

\paragraph{Experimental Setup}
We use the LBPP samples drawn in \autoref{sec:evaluation}, use \probgen with Gemini Flash 2.0, and set $N_P=10$.
\autoref{tab:ssc results} shows the pass@1 when using SSC.
In particular, we compute pass@1 by using the LBPP-provided unit tests as ground truth.
For SSC methods (PBT, \probgen) we measure pass@1 by computing the pass rate of items in the larger cluster.
If there is a tie, we average the pass rate of the items of all tied clusters.

\begin{table}[]
    \centering
    \caption{Pass@1 for different methods of semantic self-consistency (SSC).}
    \label{tab:ssc results}
    \begin{tabular}{lr} \toprule
    SSC Method   & pass@1 \\ \midrule
    None -- Baseline & 40.6 \\
    PBT &  40.6\\
    \probgen & \textbf{44.6}\\ 
    \probgen w/o Filtering & 40.6\\\bottomrule
    \end{tabular}
    
\end{table}

\paragraph{Results} The results show a mild improvement of 10\% relative to the baseline when using \probgen.
Removing the spurious counterexample filtering negates this effect.
While the improvement of \probgen is non-trivial, it is also relatively small.
This suggests that the errors the LLMs make are ``uniform'' in the space of the semantics rather than having a large cluster with the correct semantics and many small ones with rare errors.

\section{Related Work}

(Dis)proving the functional equivalence of two pieces of code is known to be generally undecidable, but in many real-life code possible. There has been a large corpus of work in the programming languages research that investigates this problem which we omit here.
A substantial number of works has focused on the ``equivalence modulo inputs'' paradigm~\citep{le2014compiler} focusing on different implementations of some standard, such as compilers, or programs that perform code transformations~\citep{daniel2007automated}.
These methods use randomized equivalence testing and predate LLMs.
\citet{kommrusch2023self} was probably the first ML work that aimed to assist equivalence proofs of straight-line programs by suggesting rewrite rules.

More recently, research has looked into test generation, such as the work of \citet{lemieux2023codamosa}. CodeT~\citep{chen2022codet} generates tests as a means of exposing the functional differences of a set of candidate implementations, similar to semantic clustering.
As discussed in \autoref{sec:probe gen}, unit test generation is a special form of probe generation that requires providing both input usage of an interface, but also predicting the expected output.
This makes it a harder task for LLMs while also suffering from spurious counterexamples~\citep{alshahwan2024automated}.
\probgen circumvents this issue by comparing the output values of a probes that only provide inputs to the tested code, with the downside that a ground truth implementation needs to be present.

\section{Conclusion}
We presented \probgen, a useful technique for disproving the equivalence of two or more programs.
\probgen has applications in the evaluation of LLM-generated or edited code in a variety of scenarios where two (or more) implementations exist.
\probgen can help strengthen the evaluation of program synthesis and expand into cases where no unit tests exists.
At the same time, \probgen represents an interesting standalone task that tests the code understanding capabilities of LLMs and future work may opt to study it in isolation.

\section*{Acknowledgements}
We would like to thank Disha Shrivastava for helpful comments on a previous draft.


\bibliography{main}
\bibliographystyle{icml2025}

\newpage
\appendix
\onecolumn

\section{Generating LBPP Implementations}
\label{appx:lbpp implementations}
We ask LLMs to generate (possibly incorrect) solutions to the LBPP problems using the following prompt template.

\begin{lstlisting}
You are {adjective} {profession} who is writing Python code. You must respond with complete Python code that does *NOT* contain any TODO comments and does not raise `NotImeplementedError` or be a plain `pass` statement. The code must be correct and complete to the best of your abilities. Include import statements when needed. The code must be included within Markdown code blocks. You MUST generate a function with the exact name in the user instruction.
\end{lstlisting}
Where the two template variables are taken from the following lists:
\begin{lstlisting}
PROFESSIONS = [
    "C++ software engineer",  "Rust software engineer", "bioinformatician", "data analyst", "data engineer", "data scientist",
    "developer", "distinguished engineer", "educator", "high-school teacher", "professor", "research scientist",
    "researcher", "software engineer", "systems developer", "teacher", "web developer",
]

ADJECTIVES = [
    "average", "beginner", "busy", "careless", "experienced", "expert", "extremely skilled", "highly skilled", "incompetent",
    "newbie", "novice", "professional", "rushed", "skilled", "talented", "tired",
]
\end{lstlisting}

\section{Success Probability per Search Strategy}
\begin{figure*}[t]
    \centering
    \begin{subfigure}{0.32\textwidth}
        \centering
        \includegraphics[width=\textwidth,trim={1cm 0.1cm 1cm 1cm},clip]{figs/flash-full.pdf}
        \caption{Lorem ipsum}
    \end{subfigure}
    \hfill
    \begin{subfigure}{0.32\textwidth}
        \centering
        \includegraphics[width=\textwidth,trim={1cm 0.1cm 1cm 1.0cm},clip]{figs/flash-dec-k-by-one.pdf}
        \caption{Lorem ipsum}
    \end{subfigure}
    \hfill
    \begin{subfigure}{0.32\textwidth}
        \centering
        \includegraphics[width=\textwidth,trim={1cm 0.1cm 1cm 1.0cm},clip]{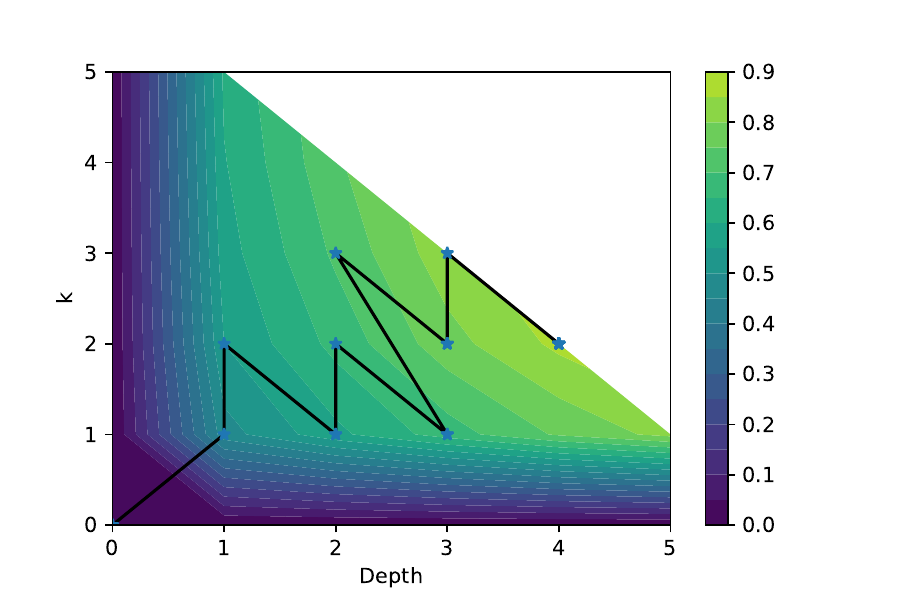}
        \caption{Lorem ipsum}
    \end{subfigure}
    
    \begin{subfigure}{0.32\textwidth}
        \centering
        \includegraphics[width=\textwidth,trim={1cm 0.1cm 1cm 1cm},clip]{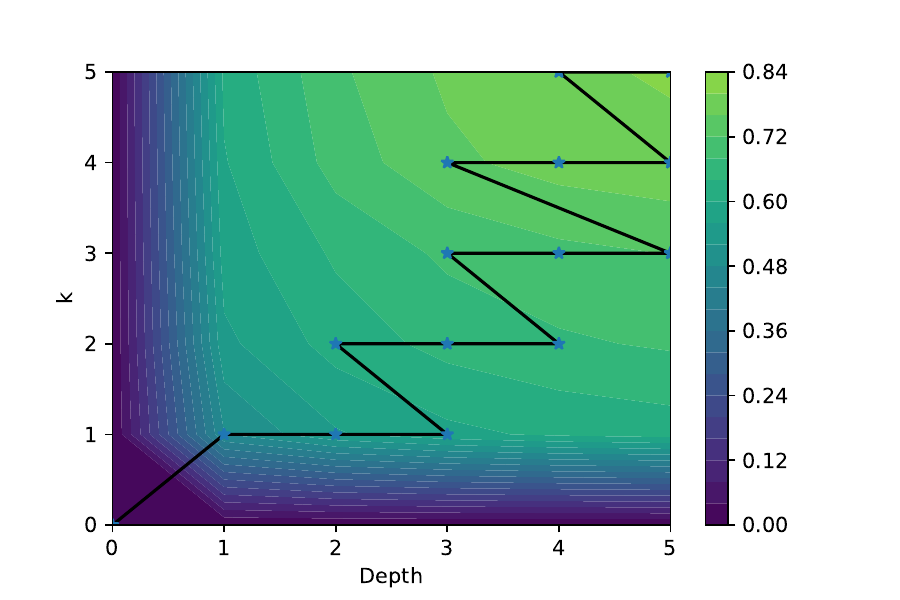}
        \caption{Lorem ipsum}
    \end{subfigure}
    \hfill
    \begin{subfigure}{0.32\textwidth}
        \centering
        \includegraphics[width=\textwidth,trim={1cm 0.1cm 1cm 1.0cm},clip]{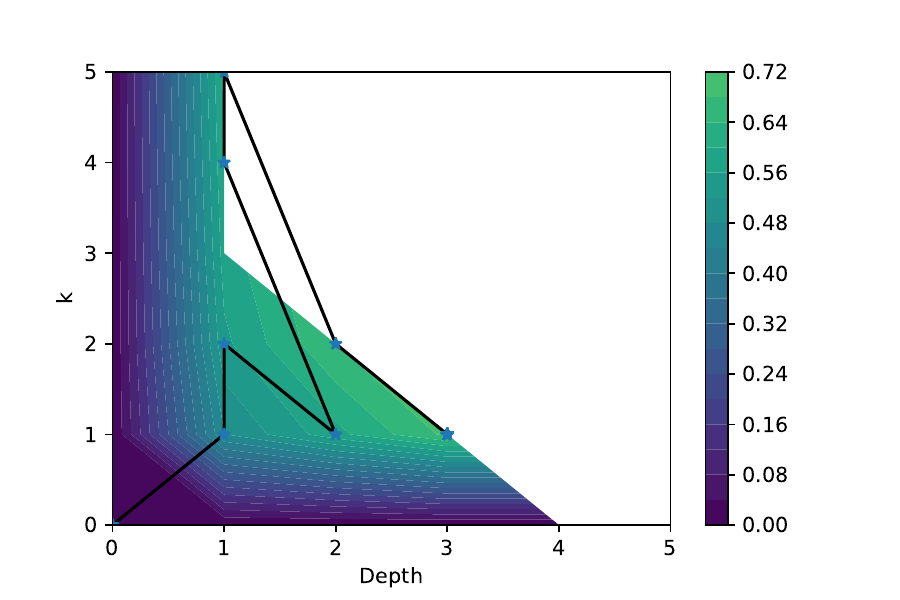}
        \caption{Lorem ipsum}
    \end{subfigure}
    \hfill
    \begin{subfigure}{0.32\textwidth}
        \centering
        \includegraphics[width=\textwidth,trim={1cm 0.1cm 1cm 1.0cm},clip]{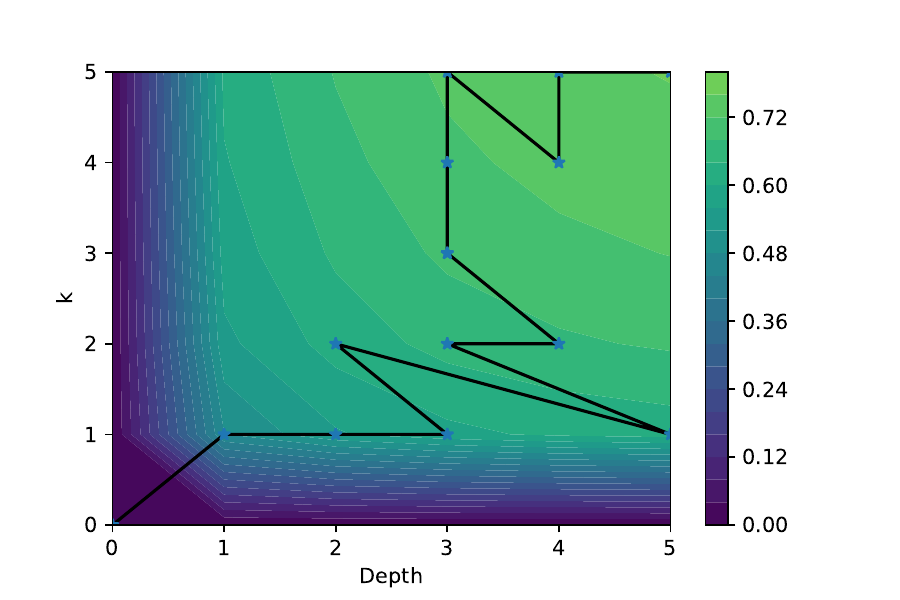}
        \caption{Lorem ipsum}
    \end{subfigure}
    \caption{Caption}
    \label{fig:search trees all}
\end{figure*}

\section{Filtering Spurious Counterexamples}
\label{appx:fp filtering prompt}
We utilize the following prompt template to filter spurious probes.
\begin{lstlisting}
Your task is to judge if the behavior of two different implementations of code are significantly different.
You will be given the following:
* A problem definition
* The input arguments
* The two different outputs returned by the two implementations.

Differences may be irrelevant if:
* The input arguments violate the (implicit) preconditions set in the problem definitions. This includes preconditions on the types and values of the inputs.
* When preconditions are violated, the way errors that are handled is irrelevant.
* The outputs differ in insignificant ways such as very minor floating-point differences or the output differs in ways that are unimportant or irrelevant to the problem definition.

Response:
* First, reason step-by-step about the way you reason about the problem.
* Then, in a new line write "RATIONALE:" followed by a very short phrase explaining your decision (e.g., "precondition violation").
* Finally, in a new line write "DIFFERENCES: " following by your "IMPORTANT" or "IRRELEVANT" decision.

### Problem

Definition
```
{query.problem_definition}
```

### Inputs
```
{query.inputs}
```

### Outputs

##### Implementation A
```
{query.output_1}
```

##### Implementation B
```
{query.output_2}
```
\end{lstlisting}

\section{Sample \probgen trajectories}
\label{appx:samples}
In these sample trajectories, \probgen found a counterexample probe, but the unit tests passed.

\begin{tcolorbox}[breakable,title=Sample 1 (Task: LBPP/78)]
\begin{tcolorbox}[colback=red!5!white,colframe=red!75!black,title=Task]
\subsection*{Task Description}

A map is represented by 1’s (for holes) and 0’s (for regular cells). A character starts at the top left cell and has to read
instructions to hopefully walk up to the bottom right cell without falling into a hole. The instructions are an ordered list of
directions (“left”, “top”, “right”, “bottom”). For each direction, the character can walk 1 or more steps in that direction.
Write a python function that takes as input the map and a list of directions and check if there is at least one path following
the instructions where the character ends at the bottom right cell without falling into a hole. Write it in Python.

\paragraph{Target function} \lstinline{isPathToBottomRightCell}

\subsection*{Version 1}
\begin{lstlisting}
<ground truth solution omitted. See task LBPP/78>
\end{lstlisting}

\subsection*{Version 2 (reformated from original to fit page)}
\begin{lstlisting}[]
def isPathToBottomRightCell(map: list, instructions: tuple) -> bool:
  rows, cols = len(map), len(map[0])
    
  def is_valid(row, col):
    return 0 <= row < rows and 0 <= col < cols and map[row][col] == 0
    
  def explore(row, col, instruction_index):
    if row == rows - 1 and col == cols - 1: return True
    
    if instruction_index == len(instructions): return False
    
    direction = instructions[instruction_index]
    if direction == "left":
        for i in range(col, -1, -1):
            if not is_valid(row, i): break
            if explore(row, i, instruction_index + 1):
                return True
    elif direction == "top":
        for i in range(row, -1, -1):
            if not is_valid(i, col): break
            if explore(i, col, instruction_index + 1):
                return True
    elif direction == "right":
        for i in range(col, cols):
            if not is_valid(row, i): break
            if explore(row, i, instruction_index + 1):
                return True
    elif direction == "bottom":
        for i in range(row, rows):
            if not is_valid(i, col): break
            if explore(i, col, instruction_index + 1):
                return True
    return False
  return explore(0, 0, 0)
\end{lstlisting}
\end{tcolorbox}
\begin{tcolorbox}[breakable,colback=blue!5!white,colframe=blue!75!black,title=LLM (Turn 1)]
The two versions implement the same logic using Depth First Search, but they differ in how they iterate through the directions and cells. The first version uses a dfs helper function with a \lstinline{currDirection} argument being incremented, and also calculates \lstinline{maxRight} and \lstinline{maxBottom} inside the function at every level. The second version uses a nested helper function explore and a \lstinline{is_valid} function, and iterates using a for loop using range. These differences are not semantically important for the result of the function. I will create an input where a path exists and the function should return \lstinline{True}, and another input where a path does not exist and the function should return \lstinline{False}.

\begin{lstlisting}
def create_fn_inputs():
    yield {
        "args": [
            [
                [0, 0, 0],
                [1, 1, 0],
                [0, 0, 0],
            ],
            ("bottom", "bottom", "right", "right"),
        ],
        "kwargs": {},
    }
    yield {
        "args": [
            [
                [0, 1, 0],
                [0, 1, 0],
                [0, 0, 0],
            ],
            ("bottom", "bottom", "right", "right"),
        ],
        "kwargs": {},
    }
\end{lstlisting}
\end{tcolorbox}

\begin{tcolorbox}[breakable,colback=green!5!white,colframe=green!75!black,title=Feedback (Turn 1)]
All inputs generated from \lstinline{create_fn_inputs()} yielded the same behavior for target function \lstinline{isPathToBottomRightCell}:

All outputs equivalent: \lstinline{True}

\begin{itemize}
    \item For input \lstinline{isPathToBottomRightCell([[0, 0, 0], [1, 1, 0], [0, 0, 0]], ('bottom', 'bottom', 'right', 'right'))} the outputs were equal: \lstinline{False}
    \item For input \lstinline{isPathToBottomRightCell([[0, 1, 0], [0, 1, 0], [0, 0, 0]], ('bottom', 'bottom', 'right', 'right'))} the outputs were equal: \lstinline{True}
\end{itemize}

Try again to generate different inputs.
\end{tcolorbox}

\begin{tcolorbox}[breakable,colback=blue!5!white,colframe=blue!75!black,title=LLM (Turn 2)]
My previous attempt failed because I made an error in the expected output for the provided inputs. The first input should have returned \lstinline{True} as a path exists using that instruction set and the second input should have returned \lstinline{True} as well since there is a valid path using the provided instructions. Both versions correctly returned True for first input and \lstinline{True} for the second input. The issue wasn't that the functions behaved differently. It was that I miscalculated if a path existed.

The difference between the two functions lies in the way they explore the map. The first version, uses recursion and increments the instruction index at each recursive call, while in the second version the recursive function \lstinline{explore} is called inside the for loop itself when moving along the map according to the provided directions. This might be a good place to make the two functions behave differently, by providing a set of instructions that, when interpreted in a different order, might lead to different results. Specifically, imagine a case where there is an early right instruction, but only a bottom path is valid. The dfs in the first version might explore the sub-branch of the \lstinline{right} path and return \lstinline{False} too early in the recursive stack. Let's construct an example where going \lstinline{right} then \lstinline{bottom} is a dead-end but going \lstinline{bottom} then \lstinline{right} find the end of the map.

\begin{lstlisting}
def create_fn_inputs():
    yield {
        "args": [
            [
                [0, 0, 1],
                [0, 1, 0],
                [0, 0, 0],
            ],
            ("right", "bottom", "bottom", "right"),
        ],
        "kwargs": {},
    }
    yield {
        "args": [
            [
                [0, 0, 0],
                [0, 1, 0],
                [0, 0, 0],
            ],
            ("right", "bottom", "bottom", "right"),
        ],
        "kwargs": {},
    }    
\end{lstlisting}
\end{tcolorbox}

\begin{tcolorbox}[breakable,colback=green!5!white,colframe=green!75!black,title=Feedback (Turn 2)]
Differentiating inputs found!

\subsection*{Differentiating inputs}
For input \lstinline{isPathToBottomRightCell([[0, 0, 1], [0, 1, 0], [0, 0, 0]], ('right', 'bottom', 'bottom', 'right')) } the outputs were different: \lstinline{False} \vs \lstinline{True}

\subsection*{Non-differentiating inputs}
For input \lstinline{isPathToBottomRightCell([[0, 0, 0], [0, 1, 0], [0, 0, 0]], ('right', 'bottom', 'bottom', 'right'))} the outputs were equal: \lstinline{True}
\end{tcolorbox}
\end{tcolorbox}

\begin{tcolorbox}[breakable,title=Sample 2 (Task: LBPP/34)]
\begin{tcolorbox}[colback=red!5!white,colframe=red!75!black,title=Task]
\subsection*{Task Description}
Write a python function \lstinline{def count_string(text: str, string: str) -> int} that calculates occurrences of a specified string within a given text. Make it case-insensitive and raise an error if the provided string has less than two characters. The string can overlap with itself.

\paragraph{Target function} \lstinline{count_string}

\subsection*{Version 1}
\begin{lstlisting}
<ground truth solution omitted. See task LBPP/34>
\end{lstlisting}

\subsection*{Version 2}
\begin{lstlisting}
def count_string(text: str, string: str) -> int:
    """
    Calculates occurrences of a specified string within a given text. 
    Make it case-insensitive and raise an error if the provided string has less than two characters.
    The string can overlap with itself.

    Args:
        text (str): The text to search within. 
        string (str): The string to search for.

    Returns:
        int: The number of occurrences of the string in the text. 

    Raises:
        ValueError: If the provided string has less than two characters.
    """

    if len(string) < 2:
        raise ValueError("The provided string must have at least two characters.")

    text = text.lower()
    string = string.lower()

    count = 0
    for i in range(len(text) - len(string) + 1):
        if text[i:i + len(string)] == string:
            count += 1
    return count
\end{lstlisting}
\end{tcolorbox}

\begin{tcolorbox}[breakable,colback=blue!5!white,colframe=blue!75!black,title=LLM (Turn 1)]
The two versions have a subtle difference in how they handle case insensitivity and string comparison. The first version performs the lowercasing and comparison at each step of the loop while the second version lowercases the full strings first and compares the substrings after lowercasing. This should make no difference. Thus, a simple set of non-edge case strings will be enough to find differneces among these functions. Both of them also have the same precondition about the length of string argument, which we need to respect.

\begin{lstlisting}
def create_fn_inputs():
    yield {"args": ("abababa", "aba"), "kwargs": {}}
    yield {"args": ("hello world", "lo"), "kwargs": {}}
    yield {"args": ("TESTtestTeSt", "test"), "kwargs": {}}
    yield {"args": ("AABAB", "aab"), "kwargs": {}}
    yield {"args": ("This is a test string", "is"), "kwargs": {}}
    yield {"args": ("aaaaaaaa", "aa"), "kwargs": {}}
\end{lstlisting}
\end{tcolorbox}

\begin{tcolorbox}[breakable,colback=green!5!white,colframe=green!75!black,title=Feedback (Turn 1)]
All inputs generated from \lstinline{create_fn_inputs()} yielded the same behavior for target function \lstinline{count_string}:

All outputs equivalent: \lstinline{True}

\begin{itemize}
    \item For input \lstinline{count_string('abababa', 'aba')} the outputs were equal: \lstinline{3}
    \item For input \lstinline{count_string('hello world', 'lo')} the outputs were equal: \lstinline{1}
    \item For input \lstinline{count_string('TESTtestTeSt', 'test')} the outputs were equal: \lstinline{3}
    \item For input \lstinline{count_string('AABAB', 'aab')} the outputs were equal: \lstinline{1}
    \item For input \lstinline{count_string('This is a test string', 'is'} the outputs were equal: \lstinline{2}
    \item For input\lstinline{ count_string('aaaaaaaa', 'aa')} the outputs were equal: \lstinline{7}
\end{itemize}
Try again to generate different inputs.
\end{tcolorbox}

\begin{tcolorbox}[breakable,colback=blue!5!white,colframe=blue!75!black,title=LLM (Turn 2)]
The previous inputs were not sufficient to expose any difference, which means the subtle difference in lowercasing and comparison did not affect any of the previous inputs. The difference in the original implementations is that the first lowercases on the fly inside the loop whereas the second lowercases the entire strings beforehand. This means an input such as \lstinline{text="aAaA", string="aa"} should, if there were any differences at all, expose differences. I'll also add other examples to increase the possibility of finding a difference.

\begin{lstlisting}
def create_fn_inputs():
    yield {"args": ("aAaA", "aa"), "kwargs": {}}
    yield {"args": ("ababa", "Ab"), "kwargs": {}}
    yield {"args": ("TestTest", "TeSt"), "kwargs": {}}
    yield {"args": ("hello world", "world"), "kwargs": {}}
    yield {"args": ("aaaaa", "aa"), "kwargs": {}}
    yield {"args": ("XYxyXYxy", "xy"), "kwargs": {}}
    yield {"args": ("zzZZ", "zz"), "kwargs": {}}
\end{lstlisting}
\end{tcolorbox}

\begin{tcolorbox}[breakable,colback=green!5!white,colframe=green!75!black,title=Feedback (Turn 2)]
All inputs generated from \lstinline{create_fn_inputs()} yielded the same behavior for target function \lstinline{count_string}:

All outputs equivalent: \lstinline{True}

\begin{itemize}
    \item For input \lstinline{count_string('aAaA', 'aa')} the outputs were equal: \lstinline{3}
    \item For For input \lstinline{count_string('ababa', 'Ab')} the outputs were equal: \lstinline{2}
    \item For For input \lstinline{count_string('TestTest', 'TeSt')} the outputs were equal: \lstinline{2}
    \item For For input \lstinline{count_string('hello world', 'world')} the outputs were equal: \lstinline{1}
    \item For For input \lstinline{count_string('aaaaa', 'aa')} the outputs were equal: \lstinline{4}
    \item For For input \lstinline{count_string('XYxyXYxy', 'xy')} the outputs were equal: \lstinline{4}
    \item For For input \lstinline{count_string('zzZZ', 'zz')} the outputs were equal: \lstinline{3}
\end{itemize}
Try again to generate different inputs.
\end{tcolorbox}

\begin{tcolorbox}[breakable,colback=blue!5!white,colframe=blue!75!black,title=LLM (Turn 3)]
The inputs still did not expose any differences. The core logic of both functions appears to be equivalent in terms of the final result, even with mixed case inputs. The problem statement specified that the functions should also raise an error if the length of the searched string is less than 2. I will add such a case to the input to verify that both implementations raise the error correctly. Also, I'll add an empty text search case.

\begin{lstlisting}
def create_fn_inputs():
    yield {"args": ("aAaA", "aa"), "kwargs": {}}
    yield {"args": ("ababa", "Ab"), "kwargs": {}}
    yield {"args": ("TestTest", "TeSt"), "kwargs": {}}
    yield {"args": ("hello world", "world"), "kwargs": {}}
    yield {"args": ("aaaaa", "aa"), "kwargs": {}}
    yield {"args": ("XYxyXYxy", "xy"), "kwargs": {}}
    yield {"args": ("zzZZ", "zz"), "kwargs": {}}
    yield {"args":("", "ab"), "kwargs":{}}
    yield {"args": ("abc", "a"), "kwargs": {}}
\end{lstlisting}
\end{tcolorbox}

\begin{tcolorbox}[breakable,colback=green!5!white,colframe=green!75!black,title=Feedback (Turn 3)]
All inputs generated from \lstinline{create_fn_inputs()} yielded the same behavior for target function \lstinline{count_string}:

All outputs equivalent: \lstinline{True}

\begin{itemize}
    \item For input \lstinline{count_string('aAaA', 'aa')} the outputs were equal: \lstinline{3}
    \item For input \lstinline{count_string('ababa', 'Ab')} the outputs were equal: \lstinline{2}
    \item For input \lstinline{count_string('TestTest', 'TeSt')} the outputs were equal: \lstinline{2}
    \item For input \lstinline{count_string('hello world', 'world')} the outputs were equal: \lstinline{1}
    \item For input \lstinline{count_string('aaaaa', 'aa')} the outputs were equal: \lstinline{4}
    \item For input \lstinline{count_string('XYxyXYxy', 'xy')} the outputs were equal: \lstinline{4}
    \item For input \lstinline{count_string('zzZZ', 'zz')} the outputs were equal: \lstinline{3}
    \item For input \lstinline{count_string('', 'ab')} the outputs were equal: \lstinline{0}
    \item For input \lstinline{count_string('abc', 'a')} the outputs were equal: \lstinline{ValueError('The string must have at least two characters.')}
\end{itemize}
Try again to generate different inputs.
\end{tcolorbox}

\begin{tcolorbox}[breakable,colback=blue!5!white,colframe=blue!75!black,title=LLM (Turn 4)]
The previous inputs still didn't reveal any difference in behavior beyond the ValueError which both functions raise. It seems like, despite the subtle implementation details, the logic is equivalent. I will now try to generate inputs that contain Unicode characters, since the \lstinline{.lower()} might behave differently depending on the system locale. It might be a long shot, but its worth trying.

\begin{lstlisting}
def create_fn_inputs():
    yield {"args": ("aAaA", "aa"), "kwargs": {}}
    yield {"args": ("ababa", "Ab"), "kwargs": {}}
    yield {"args": ("TestTest", "TeSt"), "kwargs": {}}
    yield {"args": ("hello world", "world"), "kwargs": {}}
    yield {"args": ("aaaaa", "aa"), "kwargs": {}}
    yield {"args": ("XYxyXYxy", "xy"), "kwargs": {}}
    yield {"args": ("zzZZ", "zz"), "kwargs": {}}
    yield {"args":("", "ab"), "kwargs":{}}
    yield {"args": ("abc", "a"), "kwargs": {}}
    yield {"args": ("(*@\ss\ss\ss\ss@*)", "(*@\ss\ss@*)"), "kwargs": {}}
    yield {"args": ("ΣΣΣΣ", "ΣΣ"), "kwargs": {}}
\end{lstlisting}
\end{tcolorbox}

\begin{tcolorbox}[breakable,colback=green!5!white,colframe=green!75!black,title=Feedback (Turn 4)]
Differentiating inputs found!

\subsection*{Differentiating inputs}
For input \lstinline{count_string('ΣΣΣΣ', 'ΣΣ')} the outputs were different: \lstinline{3} vs. \lstinline{1}


\subsection*{Non-differentiating inputs}
\begin{itemize}
    \item For input  \lstinline{count_string('aAaA', 'aa')} the outputs were equal: \lstinline{3}
    \item For input  \lstinline{count_string('ababa', 'Ab')} the outputs were equal: \lstinline{2}
    \item For input  \lstinline{count_string('TestTest', 'TeSt')} the outputs were equal: \lstinline{2}
    \item For input  \lstinline{count_string('hello world', 'world')} the outputs were equal: \lstinline{1}
    \item For input  \lstinline{count_string('aaaaa', 'aa')} the outputs were equal: \lstinline{4}
    \item For input  \lstinline{count_string('XYxyXYxy', 'xy')} the outputs were equal: \lstinline{4}
    \item For input  \lstinline{count_string('zzZZ', 'zz')} the outputs were equal: \lstinline{3}
    \item For input  \lstinline{count_string('', 'ab')} the outputs were equal: \lstinline{0}
    \item For input  \lstinline{count_string('abc', 'a')} the outputs were equal: \lstinline{ValueError('The string must have at least two characters.')}
    \item For input  \lstinline{count_string(}\ss\ss\ss\ss\lstinline{)', '(}\ss\ss\lstinline{')} the outputs were equal: \lstinline{3}
\end{itemize}
\end{tcolorbox}

\end{tcolorbox}

\end{document}